# A solid-state source of single and entangled photons at diamond SiV⁻-center transitions operating at 80K

Xin Cao[1], Jingzhong Yang[1], Tom Fandrich[1], Yiteng Zhang[1], Eddy P. Rugeramigabo[1], Benedikt Brechtken[1], Rolf J. Haug[1,2], Michael Zopf[1]*, Fei Ding[1,2]*

[1]Institut für Festkörperphysik, Leibniz Universität Hannover, Appelstraße 2, 30167, Hannover, Germany

[2]Laboratorium für Nano- und Quantenengineering, Leibniz Universität Hannover, Schneiderberg 39, 30167, Hannover, Germany

## ABSTRACT

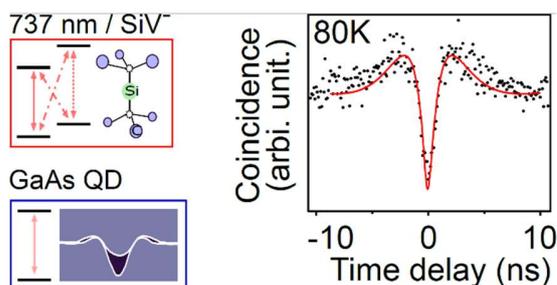

**Large-scale quantum networks require the implementation of long-lived quantum memories as stationary nodes interacting with qubits of light. Epitaxially grown quantum dots hold great potential for the on-demand generation of single and entangled photons with high purity and indistinguishability. Coupling these emitters to memories with long coherence times enables the development of hybrid nanophotonic devices incorporating the advantages of both systems. Here we report the first GaAs/AlGaAs quantum dots grown by droplet etching and nanohole infilling method, emitting single photons with a narrow wavelength distribution (736.2 ± 1.7 nm) close to the zero-phonon line of Silicon-vacancy centers. Polarization entangled photons are generated via the biexciton-exciton cascade with a fidelity of (0.73 ± 0.09). High single photon purity is maintained from 4 K ($g^{(2)}(0) = 0.07 ± 0.02$) up to 80 K ($g^{(2)}(0) = 0.11 ± 0.01$), therefore making this hybrid system technologically attractive for real-world quantum photonic applications.**

Quantum repeaters are envisioned key components for large, distributed quantum networks.[1–3] Two important ingredients for their realization are efficient single photon and entangled photon pair sources as well as platforms to store the quantum information imprinted on photons. On the one hand, epitaxially grown semiconductor quantum dots (QDs) based on GaAs have proven to be excellent single and entangled photon sources in the last years: they are on-demand emitters,[4] with ultra-high single photon purity,[5] entanglement fidelity,[6] indistinguishability,[7] and wavelength tunability.[8] However, because quantum dots are embedded in solid state matrix, the coherence time of the emitted photons are limited short due to the coupling with the phonons in the matrix.[9] On the other hand, quantum memories based on single atoms,[10] atomic ensembles[11], trapped ions[12] exhibited long storage times up to milliseconds, even though the photon generation efficiency is low. Combining QDs with a quantum memory platform allows for taking advantage of both systems by mapping quantum information encoded on 'flying' quantum bits, such as high-rate indistinguishable single and entangled photons, to stationary network nodes with adequate storage time. This would allow for realizing basic implementations of quantum repeaters via the "node receives photon" protocol.[13] However, realizing such hybrid systems is a challenging task and first of all requires exact wavelength matching between the QD photons and the storage platform. Although several groups achieved wavelength matching between InAs, InGaAs or GaAs QDs on the one hand, and trapped $Yb^+$ ions or Cs and Rb atomic vapour[14–16] on the other hand, such hybrid systems lack the compatibility with state-of-the-art semiconductor technology and therefore limit a scalable application.

Defect centers in diamond are promising solid state platforms to store and read-out single photons. The negatively charged Silicon-vacancy ($SiV^-$) and nitrogen-vacancy (NV) centers have been widely studied.[17,18] $SiV^-$ offers an efficient light-matter interface together with long spin coherence times at cryogenic temperature (13 ms[19]). Single photon storage in $SiV^-$ centers efficiently coupled with nanocavities as well as coupling to $^{13}C$ nuclear spins has been achieved, laying the foundations for implementation of practical quantum memory nodes.[20,21] The $SiV^-$ zero-phonon-line (ZPL) spectrally locates at about 737 nm,[22] within the emission regime of GaAs QDs. However, coupling experiments between $SiV^-$ centers and quantum dots have not been reported yet. To couple GaAs QDs with $SiV^-$ centers, the emission wavelength must match the ZPL. One possibility to achieve that is to grow strain-free GaAs QDs via droplet epitaxy[23] where the QD emission wavelength can be tuned by controlling the size of the metal droplets. Alternatively, *in-situ* local droplet etching can be employed to etch nanoholes, which are subsequently infilled to form the QD[24,25]. In this case, the QD emission wavelength can be precisely engineered by varying the amount of the infilled GaAs[26,27].

In this work, we present novel GaAs QDs with emission wavelengths at the $SiV^-$ ZPL, and investigate their spectral properties and single photon characteristics up to temperatures of 80 K. The GaAs QD samples are grown via molecular beam epitaxy by means of *in-situ* local droplet etching of nanoholes and subsequent infilling[28]. This technique allows for strain-free lattice matched growth of highly symmetric quantum dots that have typically been optimized for emission at 780 nm or 795 nm[29,30] to match rubidium atomic transitions. As displayed in Figure 1a, an $Al_{0.3}Ga_{0.7}As$ barrier is first grown on top of the GaAs buffer layer. Thereafter, Al droplets are deposited to etch nanoholes on the $Al_{0.3}Ga_{0.7}As$ surface. Figure 1 b and c show an atomic force microscope image and line profile of a typical nanohole, respectively. The line profile reveals nanoholes with a depth of about 11 nm and a diameter of 100 nm. The sample growth is then concluded by nanohole in-filling with GaAs and subsequently overgrowing the GaAs QDs with an $Al_{0.3}Ga_{0.7}As$ barrier layer followed by a GaAs cap layer.

Due to the much larger lateral extent of the nanohole in comparison to its depth, the energy of the confined charges and excitonic complexes is mostly determined by the confinement in the growth direction and thus the amount of in-filled GaAs. The QD emission energy is approximated by $\Delta E \propto \frac{1}{m^* d_z^2}$, where $m^*$ is the exciton effective mass and $d_z$ is the thickness of GaAs in the nanohole. Therefore, adjusting the deposited amount of GaAs allows for fine-tuning the QD emission wavelength as shown in Figure 1 d. Wavelength optimization is accompanied by micro-photoluminescence measurements performed at T = 4 K. For samples with different nominal thickness of GaAs we observe wavelength distributions centered around values ranging from approximately 700-740nm. As expected, reducing the nominal thickness of infilled GaAs leads to blue-shifting of the emission wavelength of the GaAs QDs. A typical wide-range spectrum is presented in Figure 1 e. In addition to emissions from the GaAs quantum well and AlGaAs barrier, two types of localized and spectrally well-defined emission peaks are observed at different wavelengths. The peaks highlighted in green corresponds to the emission from QDs (material filled inside the nanohole), while the blue-shifted emission close to the AlGaAs barrier emission is attributed to the residual material around nanohole borders, forming a quantum ring (QR).[31] QR formation can be explained by the droplet etching process which results in the formation of a ring-shaped structure around the nanohole (see Figure 1 b, c) because of the crystallization of residual droplet material. When GaAs is supplied after nanohole etching, most of the GaAs migrates inside the nanohole to minimize surface energy, with some GaAs material accumulating around the outer ring structure.

The optical properties of QDs can be deterministically tuned to emit at the SiV⁻ center ZPL via optimizing growth parameters such as the infilling amount of GaAs and the nanohole etching temperature. QRs can also be tuned to emit, e.g. at the NV center ZPL, however less deterministically. Therefore, the QDs are particularly attractive which is why we focus on their optical properties in this work. Here, we find that by depositing a nominal thickness of 0.56–0.6 nm of GaAs, the QDs will emit around the ZPL of SiV⁻ centers in diamond.

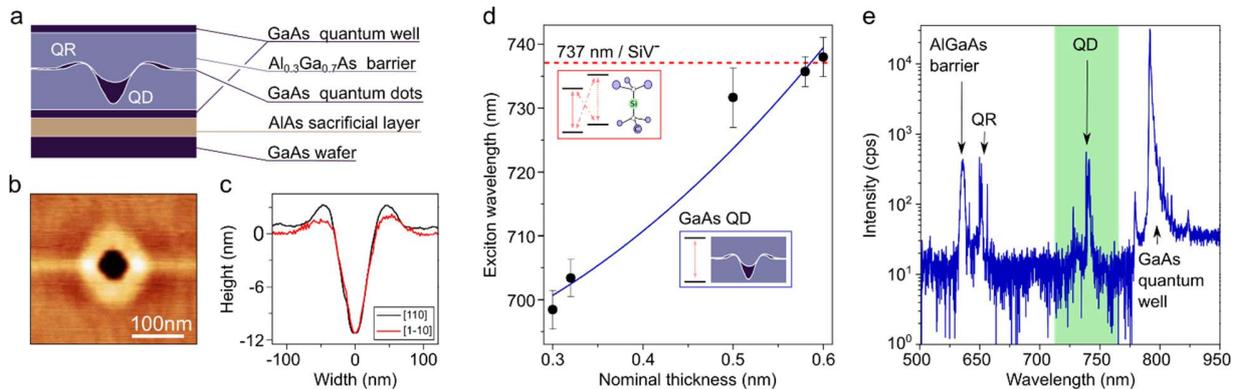

**Figure 1.** (a) Sketch of the sample structure (not to scale) with a quantum dot (QD) and quantum ring (QR). (b) Atomic force microscopy image of an Al etched nanohole on an $Al_{0.3}Ga_{0.7}As$ surface, and (c) the corresponding height profile along [110] and [1-10] direction. (d) Neutral exciton emission wavelength of the GaAs QDs as a function of deposited GaAs nominal thickness. The points are experimental data, the solid blue line is a fit and the dotted red line shows the SiV⁻ ZPL emission. (e) Micro-PL spectrum of the wafer at the location of a representative single QD emitting at around 737 nm, measured at T = 4 K.

Individual GaAs QDs are now optically excited at T = 4 K by 532 nm continuous wave laser light focused by an objective with 100x magnification and 0.7 numerical aperture. The PL spectrum of a representative QD that matches the wavelength of the SiV⁻ center ZPL is presented in Figure 2

a. The shape of the spectrum is similar to that of GaAs QDs emitting at 780-795 nm, with a dominant neutral exciton (X) emission and several red-shifted transitions typically containing the neutral biexciton (XX) and several hot trion emissions.[32] For better scalability and optimal coupling between QDs and SiV⁻ centers, it is desirable that most of the QDs on the same wafer can match the ZPL, which can be ensured by a homogeneous infilling of GaAs into the nanoholes. The left inset of Figure 2a shows the distribution of X wavelengths across the wafer, based on measurements from 32 dots and a nominal GaAs thickness of 0.56 nm. The resulting wavelength distribution is 736.2 ± 1.7 nm, matching well with the SiV⁻-ZPL. The XX peak is distinguished from the charged exciton peaks by measuring its linear polarization. The XX emission reveals the same fine structure splitting as the X emission, but with flipped linear polarization of the respective low- and high-energy spectral components (see supplementals). The coherence of the emitted X photons is measured using a Michelson interferometer and displayed in the right inset of Figure 2a, showing the interference visibility over the delay time difference between the two interferometer arms. A coherence time of 115.3 ± 5.5 ps of the X photons is extracted using a model that considers a spectral line with Gaussian broadening. This value is below the radiative lifetime-limit and points towards the presence of fast spectral diffusion due to charge traps in the QD vicinity.[33]

High single photon purity is an important property for quantum communication, e.g., to avoid photon number splitting (PNS) attacks in quantum cryptography protocols.[34] To study the purity of single photon emission from both the XX and X transitions, we perform second-order autocorrelation measurements with a standard Hanbury Brown and Twiss setup while exciting the QDs off-resonantly with a continuous wave laser. The photons are detected by avalanche photodiodes with a time resolution of 350 ps. The normalized coincidences are shown in Figure 2b together with the following theoretical model of the delay time dependent autocorrelation function [16,35,36]:

$$g^{(2)}_{XX,X}(t) = \left[(g_0 - 1)e^{\frac{-|t|}{\tau_1}} \times \left(\frac{1-\beta}{\beta}e^{\frac{-|t|}{\tau_2}}\right)\right] * IRF + 1$$

with $g_0$ the value of single photon purity, $\tau_1$ the time resulting from the pumping and decay process, $\beta$ the fraction of time when the QD is in an 'on' state, $\tau_2$ the time of the blinking process and the instrument response function *IRF*. Values of $g^{(2)}_{XX}(0)$ and $g^{(2)}_{X}(0)$ of 0.08 ± 0.01 and 0.07 ± 0.01 for X and XX are obtained, respectively, confirming a strong single photon character of the emission. The bunching effect is slightly stronger for the XX transition compared to the X transition. Because the XX peak is in spectral proximity to other charged exciton emissions (see Figure 2 a), a small fraction from these transitions might be also detected during the measurement due to imperfect spectral filtering. We attribute the slightly stronger bunching effect for the XX transition compared to the X transition to the lower probability of forming two electron-hole pairs under off-resonant excitation[37,38].

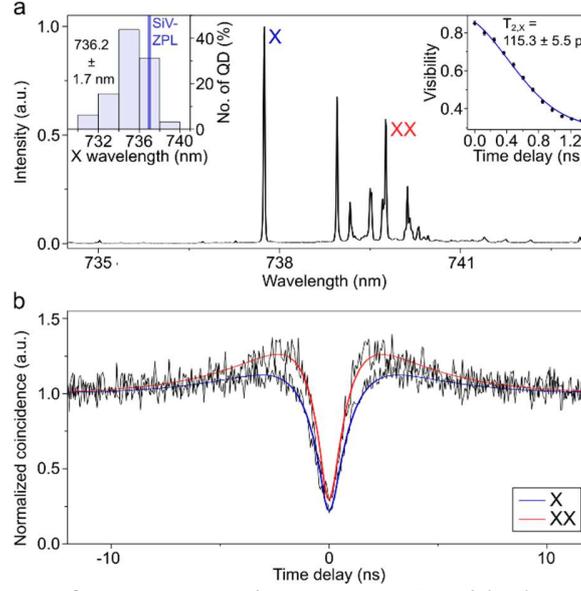

**Figure 2.** (a) PL spectrum of a representative GaAs QD with the neutral exciton wavelength matching the SiV⁻-ZPL. X and XX represent exciton and biexciton peaks, respectively. Left inset: X emission wavelength distribution obtained by measuring 32 quantum dots. Right inset: Interference visibility of the X emission over the time-delay in an un-balanced Michelson interferometer. A Gaussian lineshape is revealed and a coherence time of $T_{2,X} = 115.3 \pm 5.5$ ps determined. (b) Second-order autocorrelation measurements of X and XX emissions, respectively. The solid lines denote the theoretical model.

Polarization entanglement of multiphoton states is an important resource for quantum communication applications. The polarization state of photons emitted by a QD can be mapped to the spin state in a SiV⁻ center. If one photon from an entangled pair is used for that, polarization-spin entanglement can be achieved between the flying qubit (remaining photon) and stationary qubit (spin state in SiV⁻ center). Together with entanglement swapping, various of such systems can be used as building block for quantum repeaters to overcome the current distance limits of quantum communication.[13] Therefore, we now look at the optical properties which are relevant for generating polarization-entangled photons by exploiting the XX-X cascade. We perform a statistical measurement of the distribution of exciton fine structures splitting (FSS) in the sample which results in a value of $7.0 \pm 4.6$ µeV (see supplements). Al droplet etching has been shown to yield symmetric nanoholes on AlGaAs surfaces, which eventually result in low FSS values in QDs and therefore enable the emission of polarization-entangled XX and X photons without the need of post-growth tuning or temporal photon selection.[25,29].

Polarization-entangled photon pairs can be generated via the XX-X radiative cascade. The two-photon polarization state emitted in this decay is described by

$$|\Psi(t)\rangle = \frac{1}{\sqrt{2}}\left(|H_{XX}H_X\rangle + e^{-\frac{i}{\hbar}St_1}|V_{XX}V_X\rangle\right)$$

where S is the FSS between both bright exciton states, $t_1$ is the time the QD stays in the exciton state and $|H\rangle, |V\rangle$ are the horizontal (H) and vertical (V) polarization states. For GaAs QDs, a

maximally entangled Bell state $\Phi^+$ should be in principle measured if the fine structure splitting is 0. In our case, the FSS of the chosen QD is S = 4.9 µeV, which leads to the emission of a Bell state with a phase factor that depends on the respective exciton decay time resulting in an oscillating entanglement fidelity to one particular Bell state.[39]

Six polarization-resolved second-order cross-correlation functions are obtained and shown in Figure 3 for the rectilinear (H/V), diagonal (D/A) and circular (R/L) polarization bases together with the employed model (see supplements). Because of the angular momentum conservation, anti-bunching and bunching should ideally be observed in V-V, D-D and R-L base; while for V-H, D-A and R-R base, only anti-bunching is expected. Our results differ from this expectation due to the following reasons: First, because of the additional phase in the entanglement due to FSS and $t_1$, we expect oscillations in the D-D/D-A and R-R/R-L bases. Since the oscillation period is in the same order of magnitude like the time resolution of the employed APDs of the quantum dot, quantum beat effects are washed out. The second reason is a shift between the QD symmetry axes and the axes of the measurement system. Taking these effects into account, we assume that the entangled-photon pair polarization matrix $\rho$ can be described by the mixed state $\rho = \alpha\, \rho_{pure}(t) + (1-\alpha)\rho_{class}$, where $\rho_{pure}(t) = |\Psi(t)\rangle\langle\Psi(t)|$ is the pure two-photon density matrix and $\rho_{class} = \frac{1}{2}\, diag(1,0,0,1)$ the density matrix of classical correlated photons. By simultaneously modelling all six datasets, an entanglement fidelity of 0.73 ± 0.09 and an exciton lifetime of (368 ± 21) ps are obtained. The oscillation in the two-photon polarization states can be seen in the unconvoluted model in Fig. 3, in particular in the RR and RL bases.

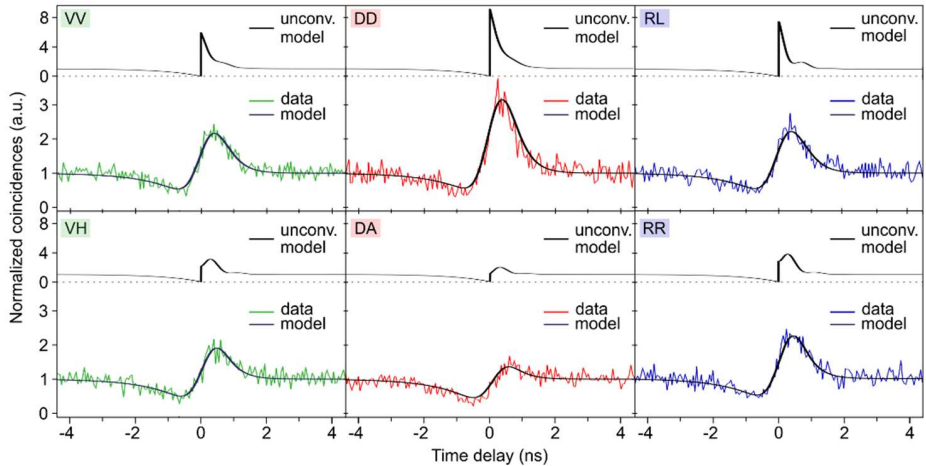

**Figure 3.** Polarization-resolved cross-correlation measurements of the XX-X emission from the biexciton-exciton cascade, with polarization projection on the H-V, D-A, and R-L base, respectively. The solid line represents the theoretical model, revealing an entanglement fidelity to the Bell state $\Phi^+$ of (0.73 ± 0.09).

The observed X lifetime is slightly longer than what is typically observed for GaAs QDs emitting at 780 nm,[29] which may be attributed to the stronger confinement due to the smaller size of the QDs here. However, the lifetime is still in a range that is expected for a system with weak carrier confinement.[40,41] Higher entanglement fidelities may be achieved by reducing the X lifetime via

Purcell enhancement in photonic nanostructures with embedded QDs. Also, post-growth tuning techniques such as anisotropic strain-tuning can reduce the FSS. Further improvements are expected when utilizing pulsed two-photon resonant excitation of the XX state and single photon detectors with better time resolution at this wavelength.

The measurements conducted so far are all performed at 4K, requiring helium to cool down the sample. Ideally, pure single photon emission should be maintained at room temperature or at least above liquid nitrogen temperature to allow for a more practical and economically feasible use of such sources. Up to now, epitaxial QDs were reported with single photon emission at room temperature or even higher temperature in material systems with a larger energy gap.[42–44] At liquid nitrogen temperatures, strained QDs have been observed to emit single photons at telecommunication wavelengths.[45,46] Strain-free GaAs/AlGaAs QDs have not yet been reported to emit single photons in the visible wavelength range at liquid nitrogen temperature. Figure 4 a shows the PL spectra of a representative QD at different temperatures. With increasing temperature, the emission peaks are red-shifted because of the shrinking band gap. Below a temperature of 40 K the phonon scattering is negligible, whereas at higher temperature it becomes more prominent, featuring an obvious phonon wing. Nevertheless, the broadened neutral X peak is still clearly visible at 80 K, above liquid nitrogen temperature. We attribute this effect to stronger quantum confinement, leading to increased electron-hole interaction and level spacing. The integrated intensity of the neutral X peak is displayed as a function of inverse temperature in Figure 4 b (solid dots and green curve). The Arrhenius function is used as a model to the data:

$$I(T) = \frac{I_0}{1 + A exp\left(\frac{-E}{k_B T}\right)},$$

where $I_0$ is the initial intensity, A is the coupling constant, $k_B$ is the Boltzmann constant, and E is the activation energy for non-radiative decay. We extract an activation energy of 16.7 ± 1.5 meV. In GaAs QDs emitting at around 785nm, an energy difference of 13 meV between $1e^1$ and $2e^1$ shells (ground and first excited electron state in the conduction band) is reported.[40] We presume that the determined activation energy of 16.7 meV for the QD emitting at 737 nm therefore corresponds to the energy difference between $1e^1$ and $2e^1$ shells. The increased spacing of energy eigenstates is directly related to decreasing QD size. With sufficient thermal energy, the charge carrier can escape the ground state and occupy higher excited states or even reach at higher temperature the energy bands of the barrier material, resulting in PL intensity quenching of the neutral X.[47]

The linewidth after subtraction of the acoustic phonon background of the neutral X peak at different temperatures is presented in Figure 4 b (open circles and blue curve). The linewidth remains almost constant below 40 K (Spectrometer resolution limit 38 μeV) and increases with ~1/(exp(1/T)) at higher temperatures. This can be modelled by an activated behavior as follows:

$$\gamma(T) = \gamma_0 + \gamma_{ac} T + \frac{a}{\exp\left(\frac{E}{k_B T}\right) - 1},$$

where γ(T) is the linewidth at temperature T, $\gamma_0$ is the linewidth at T = 0 K, $\gamma_{ac}$ stands for acoustic phonon broadening, a is a constant, $k_B$ is the Boltzmann constant, and E is the energy for optical phonon broadening.[48] The extracted energy of the longitudinal optical phonons coupling to the exciton is 21.6 ± 1.5 meV.

High single photon purity is a prerequisite for practical applications at liquid nitrogen temperature. Since the X peak is still clearly visible at 80 K, we perform second-order autocorrelation measurements from 4 K over 40 K, 60 K and up to 80 K, respectively. As shown in Figure 4 c, the $g^{(2)}(0)$ value does not change significantly at 4 K and 40 K, and only slightly increases at 60 K. At 80 K, there is still a pure single photon emission with $g^{(2)}(0)$ of about 0.1. With increasing the temperature, the anti-bunching dip at T = 0 is getting narrower. This is because on the one hand, the effective pumping rate is higher at high temperature.[49,50] On the other hand, charge carriers can relax easier between different energy states in the quantum dot, since the phonon bottleneck is overcome easier at higher temperature.[51] At 60 K and 80 K, the bunching effect in the vicinity of zero time delay is more prominent. This may be explained by the tunneling of charge carriers in and out of impurity induced trap states nearby the QD.[36] At high temperature, such processes may become more pronounced due to the increased thermal energy.

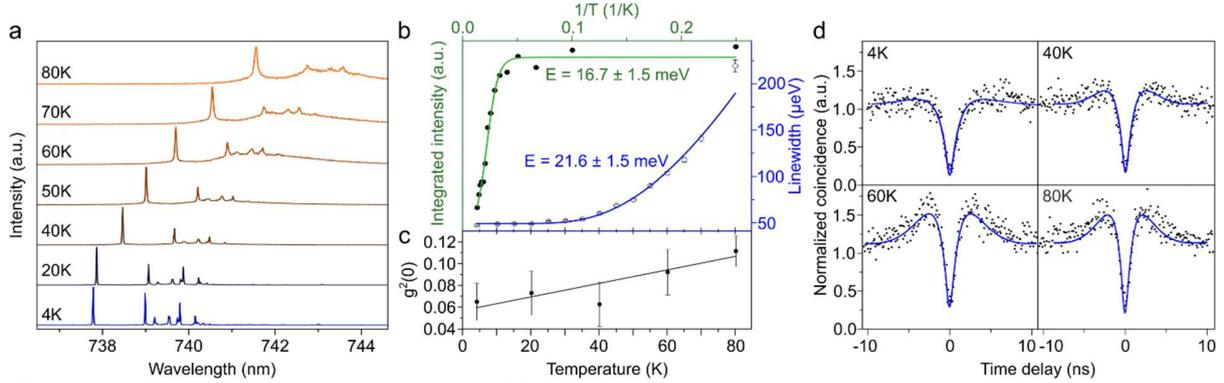

**Figure 4.** (a) Temperature dependent PL spectra showing a red-shift and emission broadening towards higher temperatures. The X emission is clearly observable even at liquid $N_2$ temperatures. (b) Integrated X peak intensity as a function of inverse temperature. (filled circles and green curve) and X linewidth as a function of temperature. (open circles and blue curve) (c) Values of $g^{(2)}(0)$ over different temperatures. (d) Second-order autocorrelation measurements of X emission at different temperatures (4 K, 40 K, 60 K and 80 K).

In conclusion, we have grown GaAs QDs and finely adjusted the infilling amount to match the ZPL of SiV$^-$ centers in diamond. Nanoholes of homogeneous depth were formed during Al droplet etching, which together with the deposition of nominally 0.56 nm GaAs results in a narrow wavelength distribution of (736.2 ± 1.7) nm. Both neutral X and XX peaks are clearly distinguished either by measuring their linear polarization state or the second-order cross-correlation functions. Both X and XX emit pure single photons, which is a prerequisite for quantum information processing. The entanglement fidelity above the classical limit of 0.5 shows that these GaAs QDs are polarization-entangled light sources with a strong potential for coupling to SiV$^-$ center based quantum memories. The slightly longer observed X lifetime as compared to GaAs QDs emitting at 780nm can prove beneficial for coupling to SiV$^-$ colour centers in diamond, which typically exhibit decay times of 1.7 ns.[52] Given that Fourier-transform limited linewidths can be reached in the QD system, the bandwidths would have a reasonably small mismatch (ratio of bandwidths of <5). By slightly modifying the photonic density of states around the quantum dot using photonic structures, an efficient coupling can therefore be envisioned. A further advantage of the presented

QD based single photon emitter is the operability above liquid nitrogen temperature, enhancing its practical feasibility.

ASSOCIATED CONTENT

**Supporting Information**.

The supplemental material (PDF) contains information about polarization dependent photoluminescence measurements and statistical measurements on the exciton fine structure. Furthermore, it contains detailed explanations about the employed model for the polarization-resolved cross-correlation measurements.

AUTHOR INFORMATION

**Corresponding Authors**
*E-mail: michael.zopf@fkp.uni-hannover.de.

*E-mail: fei.ding@fkp.uni-hannover.de.

**Author Contributions**
Fei Ding conceived the project. Michael Zopf and Eddy Rugeramigabo supervised the experiments. Xin Cao designed and grew the samples with support from Yiteng Zhang. Optical measurements were conducted by Xin Cao, Yiteng Zhang and Jingzhong Yang. Tom Fandrich developed the model for the cross-correlation measurements and analyzed the data. Surface characterization was performed by Benedikt Brechtken with support from Rolf Haug. The manuscript was written by Xin Cao, Eddy Rugeramigabo and Michael Zopf, with the contribution from all co-authors.

**Notes**
The authors declare no competing financial interest.

ACKNOWLEDGMENTS
The authors gratefully acknowledge the German Federal Ministry of Education and Research (BMBF) within the projects QR.X (16KISQ015) and SemIQON (13N16291), the European Research Council (QD-NOMS - No. GA715770, MiNet – No. GA101043851), and the Deutsche Forschungsgemeinschaft (DFG, German Research Foundation) under Germany's Excellence Strategy (EXC-2123) Quantum Frontiers (390837967). Y. Zhang acknowledges China Scholarship Council (CSC201908370225).

We thank Zhao An, Frederik Benthin and Pengji Li for fruitful discussions.